# On the Effects of Regional Spelling Conventions in Retrieval Models

Andreas Chari
a.chari.1@research.gla.ac.uk
University of Glasgow
Glasgow, United Kingdom

Sean MacAvaney
sean.macavaney@glasgow.ac.uk
University of Glasgow
Glasgow, United Kingdom

Iadh Ounis
iadh.ounis@glasgow.ac.uk
University of Glasgow
Glasgow, United Kingdom

## ABSTRACT

One advantage of neural ranking models is that they are meant to generalise well in situations of synonymity i.e. where two words have similar or identical meanings. In this paper, we investigate and quantify how well various ranking models perform in a clear-cut case of synonymity: when words are simply expressed in different surface forms due to regional differences in spelling conventions (e.g., *color* vs *colour*). We first explore the prevalence of American and British English spelling conventions in datasets used for the pre-training, training and evaluation of neural retrieval methods, and find that American spelling conventions are far more prevalent. Despite these biases in the training data, we find that retrieval models often generalise well in this case of synonymity. We explore the effect of document spelling normalisation in retrieval and observe that all models are affected by normalising the document's spelling. While they all experience a drop in performance when normalised to a different spelling convention than that of the query, we observe varied behaviour when the document is normalised to share the query spelling convention: lexical models show improvements, dense retrievers remain unaffected, and re-rankers exhibit contradictory behaviour.

https://github.com/andreaschari/regional_spelling

## 1 INTRODUCTION

Neural language models are supposed to generalise well in situations of synonymity. Whilst there have been studies on the effects of synonymity on neural models such as [5], we are not aware of any study focusing on a specific case of synonymity where synonymous words are expressed in different spelling forms due to regional language conventions — such as spelling differences between English spelling conventions. Search systems that are not robust to these surface-level differences may provide sub-optimal results when relevant content is only expressed in a specific convention, or provide disparate (unfair) exposure to content when relevant content is available in multiple conventions. Therefore, we conduct a study to answer the following research questions: **RQ1**: Do English datasets have a bias toward a specific English spelling convention? **RQ2**: Do retrieval approaches generalise across spelling convention differences in English? **RQ3**: Does normalising the spelling conventions of documents to match the query improve retrieval performance? and **RQ4**: Do neural models give query document pairs higher relevance scores when they share the same spelling convention? While some regions use combinations of spelling conventions (e.g., Indian English, Canadian English, Australian English, etc), for simplicity, we focus on only two common English spelling conventions: British and American English.

We begin our study by exploring whether MSMARCO [10] (a dataset used for training and evaluating IR systems) and C4 [13] (a dataset used for language model pre-training) exhibit a spelling convention bias, and find that they both show a clear preference towards American English spelling conventions. Based on this observation, we investigate whether retrieval systems may learn this bias during training. We test the effectiveness of a variety of lexical, dense, and neural re-ranking retrieval approaches on subsets of queries that contain either British or American spelling conventions. We observe that retrieval methods are in general robust and do not exhibit a bias on either British or American English. We then study the effect of normalising the documents to match the spelling convention of the queries as a possible performance improvement strategy. We observe that all models are affected by document normalisation, especially when the document is normalised to a different spelling convention than that of the query, where we observe drops in effectiveness. We also observe that when we normalise the document to share the spelling convention of the query, lexical models are improved, whereas neural re-rankers are harmed and dense retrievers show no significant change in performance. Finally, we perform a systematic probing of the neural models' scoring behaviour to gain more insights into how different pairs of query and relevant documents are ranked based on the normalisation of document and additionally explain why dense retrieval and neural re-ranking models are affected differently when the document is normalised to share the query spelling. We observe that even though all neural retrieval models have a very clear preference for the document sharing the query spelling convention over the document having a different spelling convention, they show no preference for normalising the document to share the query spelling over just using the document as-is. Interestingly, this contradicts the behaviour of neural re-rankers observed in the document normalisation experiments.

To summarise, we observe that even though there is a prevalence of American spelling in commonly used training datasets such as MSMARCO and C4, we find that retrieval approaches are robust enough to generalise. We also find that document normalisation has an impact on retrieval models' performance, demonstrating that we still need to be considerate of the document surface form when developing retrieval pipelines as well as studying more the effects of document normalisation.

## 2 RELATED WORK

Prompted by the challenge of explaining the decisions that neural ranking models make, several works have explored ways to identify and measure their behaviours and robustness. MacAvaney et al. [8] proposed a framework of diagnostic probes on the behaviour of pretrained language models when used for ad-hoc search. They conducted empirical studies that shed insights into factors that help neural models and identified potential biases such models may exhibit. Their framework uses document pair probing, i.e. samples of a query and two documents, which differ in one characteristic and the ranking scores of each pair are compared to study how the said characteristic impacts ranking. Zhuang et al. [15] investigated the effects of keyword typos in queries on pre-trained language-based models for retrieval and ranking. They used typo generators on a sample of queries from the MSMARCO dataset [10] and compared the performance of BERT-based rankers when using queries with and without the generated typos. Their experiments showed typos cause a significant drop in retrieval performance. Hupkes et al. [5] investigated whether neural networks are able to generalise compositionally and offered five different interpretations of compositionality. One of them deals with investigating whether neural models are robust enough to generalise on synonym substitutions. They test this by artificially substituting words with a synonym. Their results show that transformers-based models are robust enough and can pick up that there is a similarity between a word and its synonym. Inspired by the above studies, in this paper, we specifically investigate the effects regional spelling conventions have on both lexical and neural retrieval models.

Penha et al. [12] studied how variations of queries affect retrieval pipelines. Specifically queries that do not change the queries' semantics. They created a taxonomy of user created query variations such as generalisation/specialisation, aspect change, misspelling, naturality, ordering and paraphrasing with only the last four groups change the query but not its meaning. Following this taxonomy they created generators which they used to conduct experiments on a variety of ranking approaches. In addition they studied how retrieval pipelines are affected by a combination of these query variations using rank fusion. Their findings showed that each of the four variations affects retrieval performance in different extends and also that models of a similar type make similar types of error.

## 3 METHODOLOGY

For our investigation into whether ranking datasets exhibit biases towards specific English spelling conventions, we count the occurrences of convention-specific spelling, e.g., "aeroplane" for British English compared to "airplane" for American English as well as spelling convention specific terms, e.g. "barrister" for British English and "eraser" for American English using the Breame Python package [1] in both MSMARCO and C4 documents.

For the rest of the study, we focus on the subsets of MSMARCO dev queries that only contain convention-specific spelling which we identified using the Breame package. To study the effects different regional spelling conventions have on retrieval methods we first create two groups for each regional spelling convention. The British set has 173 queries containing only British spelling conventions. We consider two versions of this set: *British Queries* (), which contains only queries with British specific spelling, for example the query "what is *archaeological* context" or "what is the main *gases* of mars" and a version that normalises the queries to only American spelling conventions (→), for example the above queries are normalised to: "what is the main *gasses* of mars" and "what is *archeological* context". We apply the same process for the 1,672 queries containing only American spelling conventions to produce *American Queries* () and their normalised version (→). We also conduct document normalisation over the entire corpus. By default, it contains both spelling conventions (), and we explore normalising all documents to either British or American conventions (→ and →, respectively).

Inspired by MacAvaney et al. [8], we further probe the behaviour of neural ranking by proposing our own ABNIRML-style experiments. Given a query [q] that uses some regional spelling convention, and a document [p] that is relevant to it, we explore two normalisation methods: ConvSame which normalises the document to the same spelling convention as the query and ConvDiff which normalises the document to the opposite spelling convention. Using that we aim to test whether a neural ranking model [R] ranks R(q, ConvSame(p)) > R(q, ConvDiff(p)). In simpler terms, we investigate if a neural ranking model scores higher query-document pairs where the document [p] is normalised to the spelling convention of the query compared to when the document [p] is normalised to the opposite spelling convention. In addition, we perform a second set of ABNIRML-style experiments. Given a query [q] that uses some regional spelling convention, a document [p] that is relevant to it and one normalisation method: ConvSame, which normalises the document to the same spelling convention as the query, we aim to test whether a neural ranking model [R] ranks R(q, ConvSame(p)) > R(q, p). In simpler terms, we investigate whether a neural ranking model scores higher query-document pairs where the document [p] is normalised to the spelling convention of the query compared to when the document [p] is used without any normalisation.

## 4 EXPERIMENTAL SETUP & RESULTS

We use both MSMARCO and C4 to answer RQ1 while we we focus on MSMARCO to answer RQ2-RQ4. We run retrieval experiments on a lexical model (BM25[14]) and several neural models (MonoT5 [11], ELECTRA [2], TCT-ColBERT [7], TaS-B [4], ColBERT [6] and SPLADE [3]) using the PyTerrier [9] platform, using three different indices for MSMARCO: (i) a default index (🇺🇸🇬🇧), (ii) an index normalised to American spelling (🇺🇸🇬🇧→🇺🇸), (iii) an index normalised to British spelling (🇺🇸🇬🇧→🇬🇧). In particular, using the MS MARCO dev queries that have relevance assessments, grouped by spelling convention, we run Pyterrier retrieval pipelines with different combinations of the above indices and query and document normalisation to answer RQ2 and RQ3. For RQ4, we solely focus on the scoring behaviour of the neural models by probing them using pairs of un-normalised queries and relevant documents. We perform paired statistical testing using the Wilcoxon signed-rank test to test statistical differences and two one-sided t-tests (TOST) to test statistical equivalence of our results. We use ◇ and ∗ to denote statistically different and statistically equivalent results respectively from the default pipeline in one language. We use † and ‡ to denote statistically different and statistically equivalent results respectively from the second pipeline (documents normalised to match query spelling) of each model in one language.

Table 1: Retrieval results using *British Queries* comparing the effect of spelling conventions & normalisation. '→' indicates normalisation from spelling(s) X to spelling Y.

| Pipeline | Documents | Queries | MRR@10 | R@1000 |
|---|---|---|---|---|
| **BM25** | 🇺🇸🇬🇧 | 🇬🇧 | **0.1619** | **0.8728** |
| BM25 | 🇺🇸🇬🇧→🇬🇧 | 🇬🇧 | 0.1859 | ◇0.9306 |
| BM25 | 🇺🇸🇬🇧→🇬🇧 | 🇬🇧→🇺🇸 | ◇0.0666 | ◇0.5809 |
| BM25 | 🇺🇸🇬🇧→🇺🇸 | 🇬🇧→🇺🇸 | ‡0.1838 | ‡0.9191 |
| **BM25 » ELECTRA** | 🇺🇸🇬🇧 | 🇬🇧 | **0.4237** | **0.8728** |
| BM25 » ELECTRA | 🇺🇸🇬🇧→🇬🇧 | 🇬🇧 | ◇0.3881 | ◇0.9306 |
| BM25 » ELECTRA | 🇺🇸🇬🇧→🇬🇧 | 🇬🇧→🇺🇸 | ◇0.2929 | ◇0.5809 |
| BM25 » ELECTRA | 🇺🇸🇬🇧→🇺🇸 | 🇬🇧→🇺🇸 | ‡0.3998 | ‡0.9191 |
| **BM25 » monoT5** | 🇺🇸🇬🇧 | 🇬🇧 | **0.4150** | **0.8728** |
| BM25 » monoT5 | 🇺🇸🇬🇧→🇬🇧 | 🇬🇧 | ◇0.3790 | ◇0.9306 |
| BM25 » monoT5 | 🇺🇸🇬🇧→🇬🇧 | 🇬🇧→🇺🇸 | ◇0.2673 | ◇0.5809 |
| BM25 » monoT5 | 🇺🇸🇬🇧→🇺🇸 | 🇬🇧→🇺🇸 | ‡0.3743 | ‡0.9191 |
| **TCT** | 🇺🇸🇬🇧 | 🇬🇧 | **0.3291** | **0.9884** |
| TCT | 🇺🇸🇬🇧→🇬🇧 | 🇬🇧 | *0.3203 | 0.9884 |
| TCT | 🇺🇸🇬🇧→🇬🇧 | 🇬🇧→🇺🇸 | ◇0.2539 | ◇0.9249 |
| TCT | 🇺🇸🇬🇧→🇺🇸 | 🇬🇧→🇺🇸 | ‡0.3998 | ‡0.9827 |
| **TaS-B** | 🇺🇸🇬🇧 | 🇬🇧 | **0.3265** | **0.9884** |
| TaS-B | 🇺🇸🇬🇧→🇬🇧 | 🇬🇧 | *0.3114 | 0.9884 |
| TaS-B | 🇺🇸🇬🇧→🇬🇧 | 🇬🇧→🇺🇸 | ◇0.2419 | *0.9653 |
| TaS-B | 🇺🇸🇬🇧→🇺🇸 | 🇬🇧→🇺🇸 | ‡0.3254 | 0.9884 |
| **SPLADE** | 🇺🇸🇬🇧 | 🇬🇧 | **0.3734** | **0.9884** |
| SPLADE | 🇺🇸🇬🇧→🇬🇧 | 🇬🇧 | *0.3657 | 0.9884 |
| SPLADE | 🇺🇸🇬🇧→🇬🇧 | 🇬🇧→🇺🇸 | ◇0.3092 | ◇0.9538 |
| SPLADE | 🇺🇸🇬🇧→🇺🇸 | 🇬🇧→🇺🇸 | ‡0.3578 | ‡0.9827 |
| **ColBERT** | 🇺🇸🇬🇧 | 🇬🇧 | **0.3502** | **0.9855** |
| ColBERT | 🇺🇸🇬🇧→🇬🇧 | 🇬🇧 | *0.3387 | *0.9827 |
| ColBERT | 🇺🇸🇬🇧→🇬🇧 | 🇬🇧→🇺🇸 | ◇0.2620 | ◇0.8960 |
| ColBERT | 🇺🇸🇬🇧→🇺🇸 | 🇬🇧→🇺🇸 | ‡0.3276 | ‡0.9740 |

Table 2: Retrieval results using *American Queries* comparing the effect of spelling conventions & normalisation. '→' indicates normalisation from spelling(s) X to spelling Y.

| Pipeline | Documents | Queries | MRR@10 | R@1000 |
|---|---|---|---|---|
| **BM25** | 🇺🇸🇬🇧 | 🇺🇸 | **0.1887** | **0.8686** |
| BM25 | 🇺🇸🇬🇧→🇺🇸 | 🇺🇸 | 0.1922 | ◇0.8769 |
| BM25 | 🇺🇸🇬🇧→🇺🇸 | 🇺🇸→🇬🇧 | ◇0.1009 | ◇0.6445 |
| BM25 | 🇺🇸🇬🇧→🇬🇧 | 🇺🇸→🇬🇧 | ‡0.1963 | ‡0.8764 |
| **BM25 » ELECTRA** | 🇺🇸🇬🇧 | 🇺🇸 | **0.3802** | **0.8686** |
| BM25 » ELECTRA | 🇺🇸🇬🇧→🇺🇸 | 🇺🇸 | *0.3653 | *0.8767 |
| BM25 » ELECTRA | 🇺🇸🇬🇧→🇺🇸 | 🇺🇸→🇬🇧 | ◇0.3174 | ◇0.6445 |
| BM25 » ELECTRA | 🇺🇸🇬🇧→🇬🇧 | 🇺🇸→🇬🇧 | ‡0.3792 | ‡0.8764 |
| **BM25 » monoT5** | 🇺🇸🇬🇧 | 🇺🇸 | **0.3648** | **0.8686** |
| BM25 » monoT5 | 🇺🇸🇬🇧→🇺🇸 | 🇺🇸 | *0.3622 | ◇0.8767 |
| BM25 » monoT5 | 🇺🇸🇬🇧→🇺🇸 | 🇺🇸→🇬🇧 | ◇0.3065 | ◇0.6445 |
| BM25 » monoT5 | 🇺🇸🇬🇧→🇬🇧 | 🇺🇸→🇬🇧 | ‡0.3627 | ‡0.8764 |
| **TCT** | 🇺🇸🇬🇧 | 🇺🇸 | **0.2971** | **0.9692** |
| TCT | 🇺🇸🇬🇧→🇺🇸 | 🇺🇸 | *0.2969 | ◇0.9707 |
| TCT | 🇺🇸🇬🇧→🇺🇸 | 🇺🇸→🇬🇧 | *0.2600 | ◇0.9277 |
| TCT | 🇺🇸🇬🇧→🇬🇧 | 🇺🇸→🇬🇧 | ‡0.2983 | ‡0.9618 |
| **TaS-B** | 🇺🇸🇬🇧 | 🇺🇸 | **0.3098** | **0.9759** |
| TaS-B | 🇺🇸🇬🇧→🇺🇸 | 🇺🇸 | *0.3080 | *0.9765 |
| TaS-B | 🇺🇸🇬🇧→🇺🇸 | 🇺🇸→🇬🇧 | *0.2792 | *0.9507 |
| TaS-B | 🇺🇸🇬🇧→🇬🇧 | 🇺🇸→🇬🇧 | ‡0.3027 | ‡0.9722 |
| **SPLADE** | 🇺🇸🇬🇧 | 🇺🇸 | **0.3509** | **0.9834** |
| SPLADE | 🇺🇸🇬🇧→🇺🇸 | 🇺🇸 | *0.3499 | 0.9834 |
| SPLADE | 🇺🇸🇬🇧→🇺🇸 | 🇺🇸→🇬🇧 | *0.3227 | ◇0.9637 |
| SPLADE | 🇺🇸🇬🇧→🇬🇧 | 🇺🇸→🇬🇧 | ‡0.3472 | ‡0.9783 |
| **ColBERT** | 🇺🇸🇬🇧 | 🇺🇸 | **0.3147** | **0.9521** |
| ColBERT | 🇺🇸🇬🇧→🇺🇸 | 🇺🇸 | *0.3126 | *0.9530 |
| ColBERT | 🇺🇸🇬🇧→🇺🇸 | 🇺🇸→🇬🇧 | *0.2776 | ◇0.9085 |
| ColBERT | 🇺🇸🇬🇧→🇬🇧 | 🇺🇸→🇬🇧 | ‡0.3105 | ‡0.9486 |

**RQ1: Do English datasets have a bias toward a specific English spelling convention?**

For both MSMARCO and C4, we notice a clear preference for American English. For MSMARCO, we observe that 78% of the documents contain only American terms/spelling and 22% contain only British terms/spelling. Similar percentages were observed for the C4 dataset, with 70% of the counted documents having only American terms/spelling and 30% having only British terms/spelling. Based on these results, we can conclude that both MSMARCO and C4 have a preference for American English.

**RQ2: Do neural approaches generalise across spelling convention differences in English?**

Table 1 shows the results of our retrieval experiments on the British queries, while Table 2 shows the corresponding results on the American queries. We compare two pipelines: a pipeline whose documents were normalised to match the original query spelling convention and a pipeline converting both the queries and their corresponding documents to the query's opposite spelling convention. We observe that for both originally British queries and originally American queries there is no significant difference between the two pipelines for most models, with the exception of some *American English Queries* pipelines in Table 2 which exhibit higher variability. Overall, we can conclude that retrieval methods generalise well across regional spelling conventions showing their robustness in terms of this specific case of textual similarity.

**RQ3: Does normalising the spelling conventions of documents to match the query improve retrieval performance?**

From Tables 1 & 2 we observe that there is an improvement in the recall performance of BM25 when the queries and the documents share the same spelling conventions. We can also see that when normalising the document and query to opposite regional spelling conventions, there is a significant drop in performance. Curiously, there is a significant drop in the MRR performance of the neural re-rankers when the document is normalised to the same spelling convention as the query. Neural re-rankers also exhibit a significant performance drop when the queries are normalised to opposite spelling conventions. For dense retrievers, even though there is no significant performance change when normalising the documents to match the queries' regional spelling convention there is a significant drop in performance when the queries and documents are normalised to opposite regional spelling conventions.

**Table 3: Document normalisation preference by query spelling and model.**

| | Same vs opp. doc norm. | | | | Same vs no doc norm. | | | |
| --- | --- | --- | --- | --- | --- | --- | --- | --- |
| | 🇬🇧 Queries | | 🇺🇸 Queries | | 🇬🇧 Queries | | 🇺🇸 Queries | |
| Doc Norm → | 🇬🇧 | 🇺🇸 | 🇺🇸 | 🇬🇧 | 🇬🇧 | None | 🇺🇸 | None |
| monoT5 | 63% | 33% | 39% | 50% | 36% | 15% | 7% | 8% |
| ELECTRA | 89% | 7% | 80% | 8% | 35% | 9% | 5% | 2% |
| TCT | 81% | 14% | 77% | 11% | 33% | 10% | 4% | 2% |
| TaS-B | 86% | 9% | 79% | 9% | 31% | 13% | 5% | 2% |
| SPLADE | 72% | 23% | 68% | 20% | 30% | 13% | 4% | 3% |
| ColBERT | 91% | 5% | 84% | 5% | 36% | 8% | 5% | 1% |

Hence, we can conclude that while all retrieval methods are harmed by normalising the documents and queries to different regional spelling conventions all three types of retrieval models (lexical, neural re-rankers and dense retrievers) are affected in different ways by normalising the documents to match the queries' regional spelling convention, with BM25 being improved recall-wise, neural re-rankers getting their MRR performance harmed and with dense retrievers not showing any significant change. This is an unexpected observation since one would expect that if both queries and documents share the same regional spelling conventions, this would have a consistent positive behaviour in ranking. We leave the investigation of this behaviour to future work.

**RQ4: Do neural models give query document pairs higher scores when they share the same spelling convention?**

Table 3 presents the results of our ABNIRML experiments. We use the previous query groups and pair them with their relevant documents making 181 British query-qrel pairs and 1,789 American query-qrel pairs. We measure the percentage of pairs where the neural ranking model either gives a higher relevance score when the query shares the same spelling convention with the document or when having a different spelling convention. For example, in queries with British spelling conventions, if the ranker scores the pair higher when the document is normalised to British compared to when the document is normalised to American. The exact same procedure is applied for queries with American spelling conventions. For all models in both spelling conventions, with the exception of MonoT5, we observe that pairs that share the same spelling conventions tend to have higher scores. MonoT5 is a notable exception, showing that it does indeed score pairs with British queries higher if the documents share the same spelling convention but for pairs with American queries it scores the pair higher when they have mismatched spelling conventions.

We also measure the percentage of pairs where the neural ranking model either gives a higher relevance score when the document shares the same spelling convention with the query or instead prefers that the document is left without any normalisation. For example, in queries with British spelling conventions, if the ranker scores the pair higher when the document is normalised to British compared to when the document has no normalisation. The exact same procedure is applied for queries with American spelling conventions. We observe that in both British and American query-qrel pairs all the models score most pairs the same regardless of whether the document is normalised to the query spelling or left un-normalised.

To answer our research question, we conclude that neural models do indeed give query-document pairs higher relevance scores when they share spelling conventions compared to when normalised to the opposite spelling convention. On the other hand, they are clearly ambivalent whether the document is normalised to share the query spelling convention or left as-is, something which contradicts our previous observations from RQ3.

## 5 CONCLUSIONS

In this paper, we investigated how retrieval methods deal with cases of regional spelling synonymity using two different English spelling conventions: American and British English. We first studied whether the English datasets commonly used to pre-train neural retrieval methods have a preference towards a specific English regional spelling convention. We observed a clear preference for American English on MSMARCO and C4 (RQ1). We next showed that retrieval methods tend to generalise well with regional spelling synonymity (RQ2). In addition, we observed a varied behaviour in terms of the effect of normalising the document to share the spelling convention of the query with BM25 being improved, neural re-rankers harmed and dense retrievers not showing any significant change. On the other hand all models showed a consistent negative effect when the query and document are normalised to different spellings (RQ3). Finally (RQ4), we studied the ranking behaviour of neural ranking models and observed that most models, while clearly not preferring the queries and document to have different spelling conventions, they are generally ambivalent whether the document is normalised to match the spelling convention of the query or left as-is, something which is clearly contradicting our observations on neural re-rankers from RQ3. We believe this paper contributes an important sanity check on the robustness of current state-of-the-art re-ranking & dense retrieval approaches and provides further motivation to study the ranking behaviour of IR models on other textual characteristics as well as the surface form of the retrieval documents and their impact on Information Retrieval.

For future work, we aim to further investigate the behaviour of document normalisation and why are lexical methods, neural re-rankers and dense retrievers affected differently. We also would like to investigate the reasons for the higher variability of some of the *American English Queries* pipelines and also explore the effects regional spelling conventions have in other Non-English languages such as Brazilian Portuguese and European Portuguese. Finally user studies could be conducted to identify whether users prefer search systems biased towards their own regional spelling conventions.